# Quantum Dipolar Coupling Thermal Correction for NMR Signal during Natural Rock Flooding by Melding Experimentation and Numerical Simulation (Th-CENS)

Omar Alfarisi, Hongtao Zhang, Djamel Ouzzane, Hongxia Li, Aikifa Raza

**Abstract**
Researchers have used NMR to measure multi-phase fluid saturation and distribution inside porous media of natural rock. However, the NMR signal amplitude suffers reduction with the increase of temperature. The main reason is the Transverse Overhauser Effect, where heating increases the freedom for ionic motion, affecting spinning behavior by having two spins go in two opposite directions to form the Dipolar Coupling. We approach solving NMR thermal effects correction by melding experimentation and numerical simulation method. We use NMR for Cretaceous carbonate rock multi-phase flow research. We conduct time step in-situ temperature measurement for four different sections of the flooding system at the inlet, center, and outlet along the flooding path. In addition, we conduct a temperature measurement at the NMR device radial axis, representing the permanent magnet temperature. We build a 3D cylindrical heat transfer model for the numerical simulator that simulates thermal effect distribution on the NMR for optimally generating the correction model. The insight provided by the simulator improved the understanding of the thermal distribution at the natural rock core plug to produce a better thermal correction model that meld experimentation and simulation, a method we call Th-CENS.

**Introduction**
Planetary exploration on Earth, Moon, Mars, and beyond requires real-time digital rock typing [1] to make optimal decisions. Complex rock morphology like Cretaceous carbonate made rock typing [2-7] difficult for scientists. The critical determining properties for rock typing are pore throat size and network [8], capillary pressure [9-14], permeability [9, 13, 15-32], and fluid saturations [33-42], which are also the most difficult to measure in heterogeneous fabric [43]. NMR came to provide a simultaneous solution for measuring fluid saturation of multiple phases and determining a lithology-independent porosity [44-46], pore throat size and network, permeability, and capillary pressure [8, 13]. However, the NMR signal amplitude suffers reduction with the increase of temperature [47]. This attenuation in signal amplitudes is due to the Transverse Overhauser Effect [48], caused by heat attribution that increases the degree of freedom for ionic motion. The result is the change in the spinning behavior by having two spins go in two opposite directions, known as [49] "Dipolar Coupling" [50-52]. In our research, we use NMR to quantify the changes of fluid volumes inside natural rock during multi-phase flow at elevated temperatures. Therefore, we cannot avoid increasing the fluid temperature because we must mimic the geological formation temperature condition during the flooding experiment. The NMR Device of 0.5 Tesla, shown in Figure 1, has two units, the processing unit (left side unit with the black monitor on top) and the NMR/MRI acquisition device (right side white box of Figure 1). At the same time, the natural rock core plug holder is inside the flooding system box, surrounded by a large permeant magnetic field of the NMR/MRI acquisition device. We use the NMR/MRI measurement to determine the fluid saturation throughout the flooding experiment. We target a proper thermal correction model because quantum dipole coupling introduces the NMR signal as temperature-sensitive measurement tool. This paper describes the integration of experimentation and numerical simulation to deliver the Thermal-Correction melded Experiment and Numerical Simulation (Th-CENS) model.

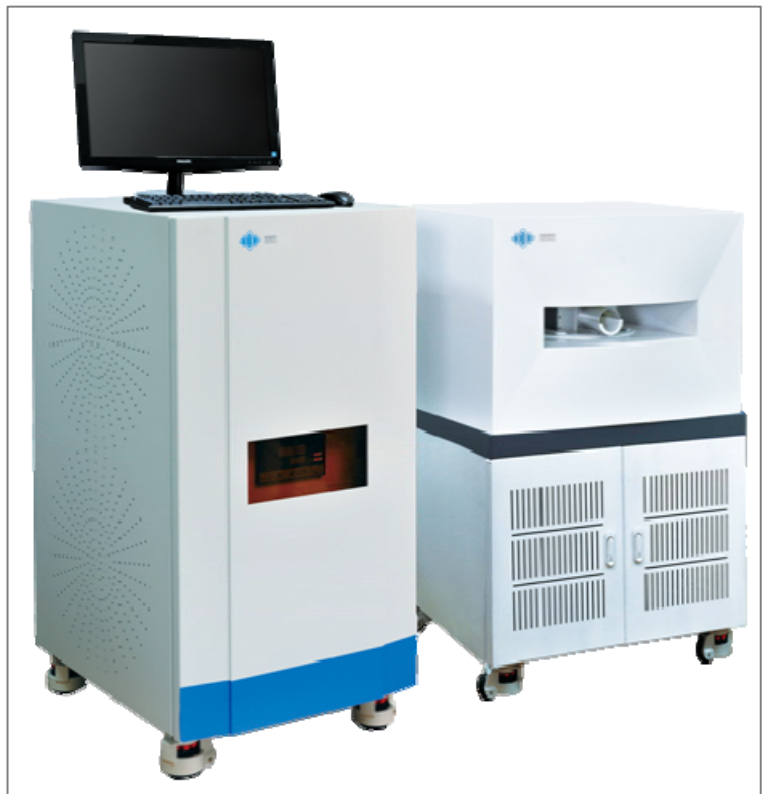

**Figure 1. Flooding Experiment NMR Device with 0.5 Tesla.**

## Method

We correct the NMR signal by conducting in-situ temperature measurements of the NMR flooding system at different locations and times. Then build a 3D numerical simulation model, and finally meld the two methods for ultimate representation and prediction of the thermal correction model. We use the Finite Element Method (FEM) to understand the temperature profile across the flooding section and the effect of ambient cooling.

1- Experimentation Setup

The NMR device is a high-pressure displacement MRI system for rock analysis. The rock core plug holder passes through the center axis of the magnetic field generated by the NMR permanent magnet. The dimensions of the rock core plug holder are in Figure 2.

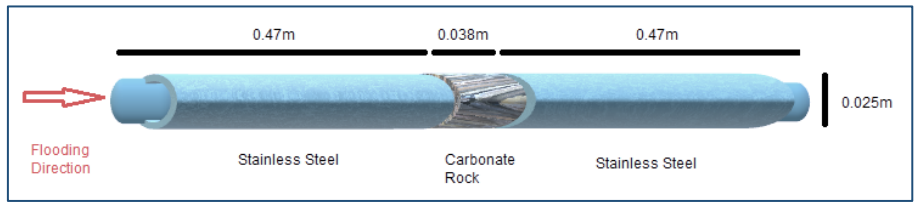
**Figure 2. The Rock Core Plug Holder (not to scale drawing).**

The cross-sectional sketch, Figure 3, shows the position of the Rock Core Assembly (enclosed by the grey dashed lines) inside the NMR device. While the Rock Core Plug Holder, Confining Fluid Chamber, and the Holder Insulator are all part of the Rock Core Assembly. In Figure 3, the red and blue rectangles represent the permanent magnet.

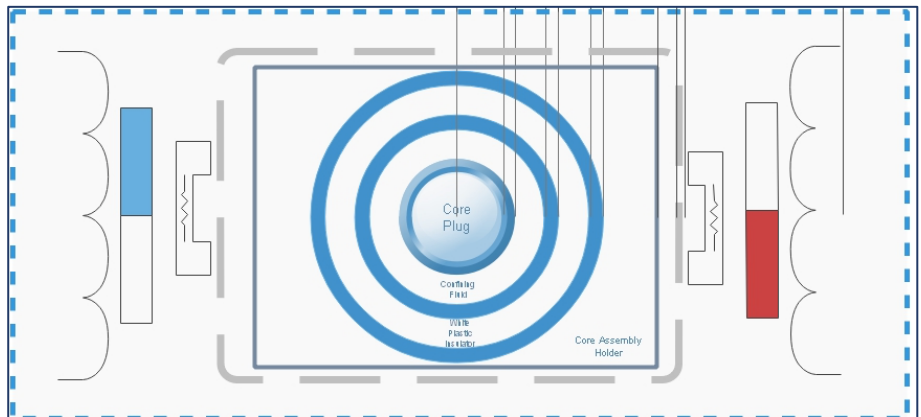
**Figure 3. NMR Device cross-section (Not-to-Scale drawing).**

2- Numerical Simulation Setup

We build an FEM model for heat transfer by replicating the cylindrical shape, Figure 4, of the rock core plug holder shown in Figure 2 by designing half of the holder, as the second half is a replica.

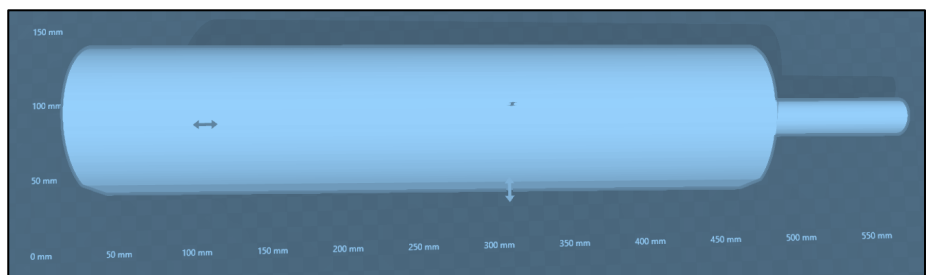
**Figure 4. 3D Design of the Rock Core Plug Holder.**

## Running Experiment and Simulation

1- Experimentation Run

Initially, we measured the laboratory and NMR nominal temperatures of 23°C and 32°C respectively, while recording the NMR signal. Then we record the NMR signal and measure the temperature again when the rock core plug holder is 29°C. The NMR signal for both measurements is in Figure 5, where the heating effect on the signal is evident.

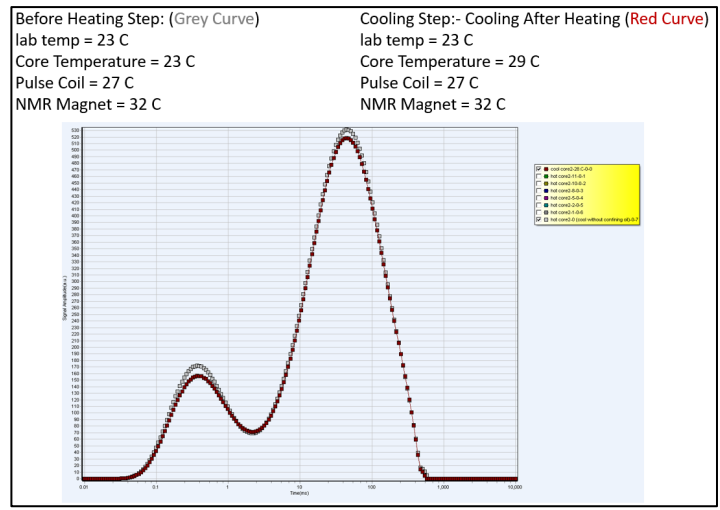

**Figure 5. NMR signal at two different rock flooding temperatures.**

Then we heat the rock core plug, shown in Figure 2, to 65°C and re-record the NMR signal to identify the change in the signal amplitude, Figure 6, noticing a more significant NMR signal drift. We continue to measure the temperature while recording NMR Signal, Table 1, to use it for the NMR signal thermal correction model by measuring the temperature at different time steps and locations.

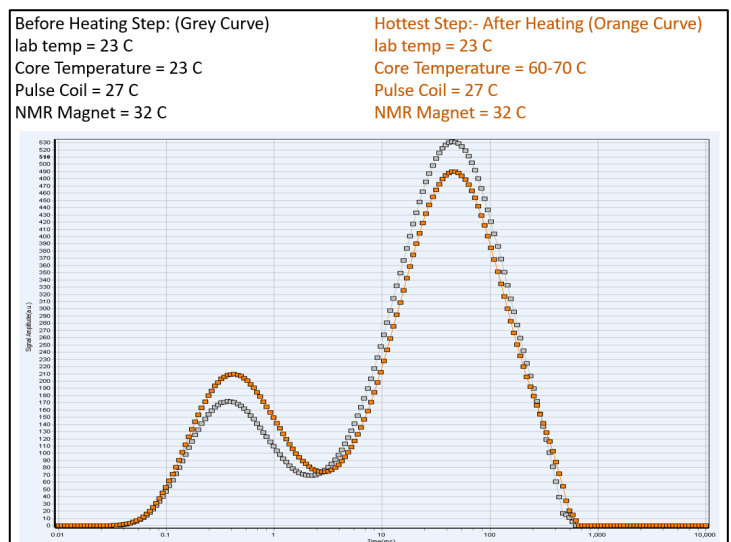

**Figure 6. NMR Signal measurement (orange color) after heating to 65°C.**

Table 1. NMR Heating/Cooling Experiment Data.

| Step Description | Curve Color | Time | Time Difference from the step above (minutes) | NMR T2 Amplitude of 1st Peak | NMR T2 Amplitude of 2nd Peek | Measured Temp. of core Assembly Frontend (°C) | Measured Temp. of core Assembly Backend (°C) | Averaged Temp. of core Assembly (°C) | Measured Rock Core Plug Temperature (°C) | Estimated Rock Core Plug Temperature from Figure 8 (°C) |
|---|---|---|---|---|---|---|---|---|---|---|
| Coldest (Lab) | Grey | 9:41 PM | | 172.3 | 531.4 | 23 | 23 | **23** | | 41 |
| Hottest | Orange | 10:03 PM | | 209.9 | 489.7 | 31 | 43 | **37** | 65 | 65 |
| Cooling 1 | Light Blue | 10:08 PM | 5 | 214.3 | 498.9 | 30 | 38 | **34** | | 60 |
| Cooling 2 | Pink | 10:22 PM | 14 | 230.4 | 524.8 | 27 | 34 | **30.5** | | 54 |
| Cooling 3 | Dark Blue | 10:42 PM | 20 | 239.4 | 538.7 | 25 | 29 | **27** | | 48 |
| Cooling 4 | Light Green | 10:50 PM | 8 | 245.7 | 541.2 | 25 | 28 | **26.5** | | 47 |
| Cooling 5 | Dark Green | 10:59 PM | 9 | 245.3 | 543.1 | 25 | 27 | **26** | 46 | 46 |
| Overnight Cooling | Red | 10:22 AM +1day | 683 | 156.5 | 518.6 | 23 | 23 | **23** | | 41 |

2- Numerical Simulation Run

We use the maximum allowable mesh size to consume less computational power. Then we increase the resolution of the mesh size to see if there is any effect of the mesh size on the simulation results. The finding showed a negligible effect, and the largest mesh size is usable without affecting the outcome. We use the largest mesh size, 30% of the core holder hole radius, Figure 7. Once the mesh is ready, we feed the simulator with stainless steel's corresponding heating coefficients [53-55] and run the simulator.

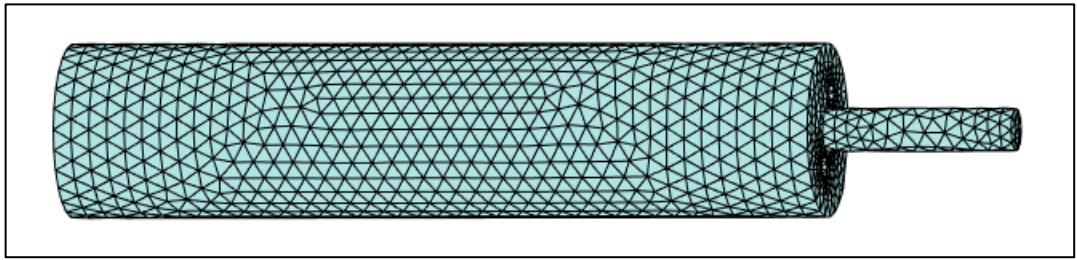

Figure 7. FEM model mesh representation of the 3D cylindrical geometry.

**Result and Discussion**

1- Experimentation

We have only two measured data points representing the temperature of the natural rock core plug, Table 1. One data point is at the hottest step with 65°C, while the 2nd point is at the 5th cooling step with 46°C. However, we need more data points to have a better thermal correction model. Therefore, we interpolate between 65°C and 46°C using the average temperature of the core assembly available for all steps, as shown in Figure 8.

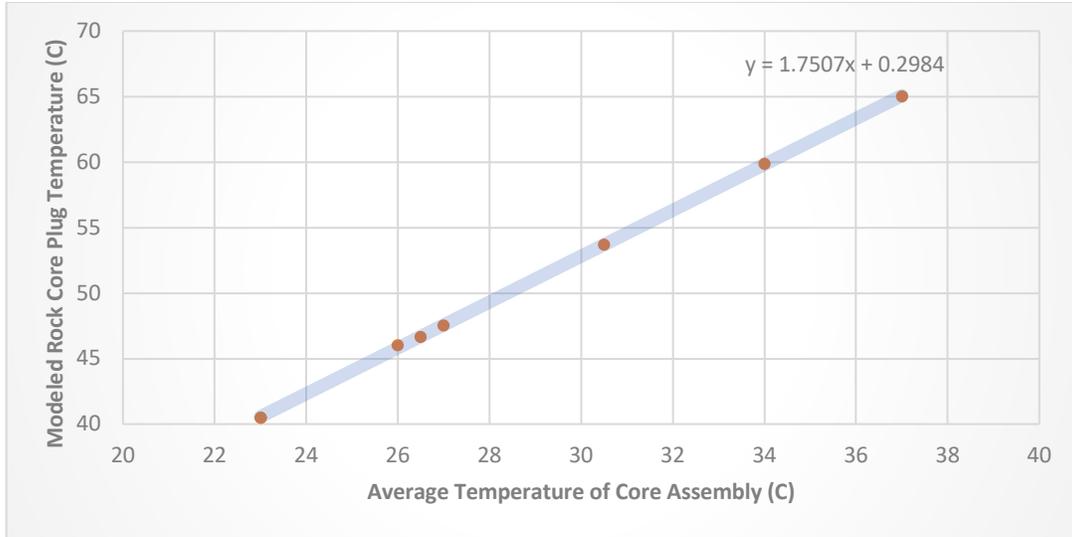

**Figure 8. Interpolating Rock Core Plug Temperature from the Average Temperature of the Core Assembly.**

Then we find the cooling rate (heat transfer) of the rock core plug with time, which forms a logarithmic function, as shown in Figure 9, and described in Eq. 1 below:

$$T_{rock_{core}} = 70.086 - 5.93 \ln(t) \tag{Eq. 1}$$

where,
$T_{rock_{core}}$ is the temperature (°C) of the rock core plug,
$t$ is time (minutes).

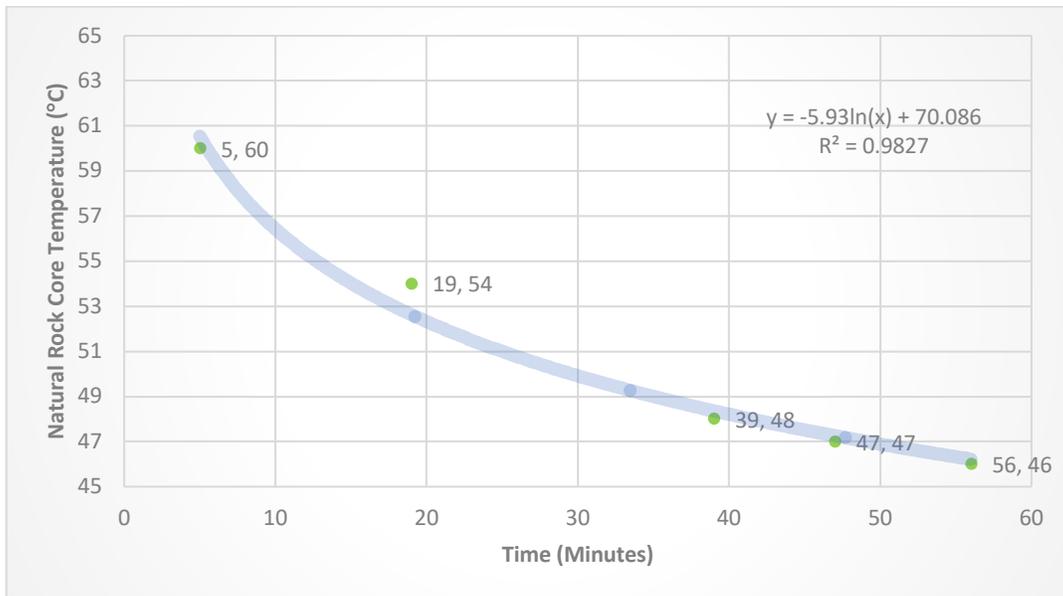

**Figure 9. Experiment results plot of the heat transfer (time-dependent) model.**

Then, we analyze the changes in the NMR T2 amplitude with the temperature of the rock core plug, as shown in Figure 10. The following correction model, demonstrated in Eq. 2, is for the NMR T2 2nd peak (the left-side peak of Figure 6):

$$A_{NMR-T2_{2nd}} = 679.98 - 2.9485\, T_{rock_{core}} \quad \text{(Eq. 2)}$$

where,

$A_{NMR-T2_{2nd}}$ is NMR T2 2nd Peak Amplitude,

$T_{rock_{core}}$ is rock core plug temperature (°C).

While the following correction model, described in Eq. 3, and shown in Figure 10, is for the NMR T2 1st peak (the right-side peak of Figure 6):

$$A_{NMR-T2_{1st}} = 336.45 - 1.9802\, T_{rock_{core}} \quad \text{(Eq. 3)}$$

where,

$A_{NMR-T2_{1st}}$ is NMR T2 1st Peak Amplitude,

$T_{rock_{core}}$ is rock core plug temperature (°C).

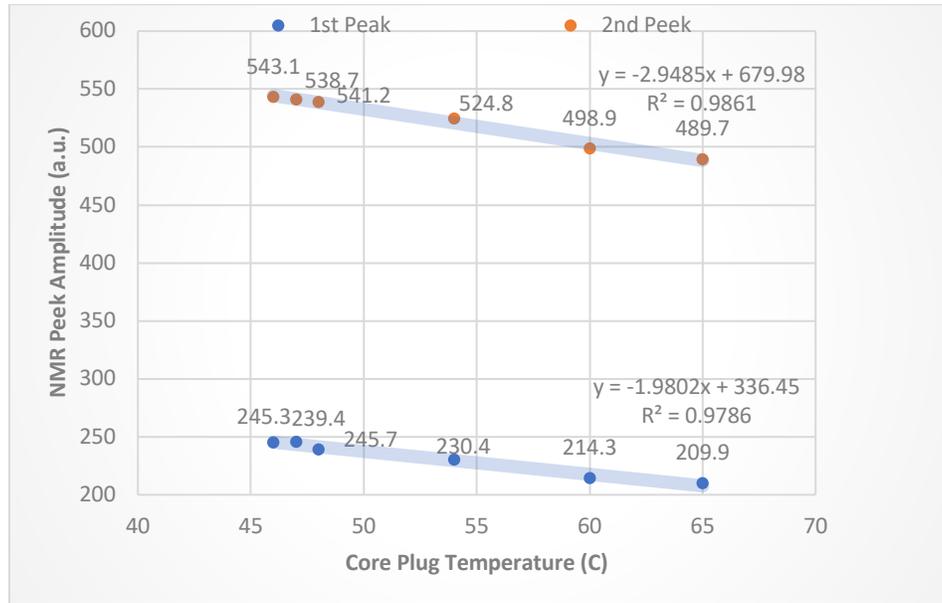

**Figure 10. NMR T2 peaks temperature correction model.**

2- Numerical Simulation

After building and configuring the FEM model, the first run suffered an extended processing time when using a high-resolution mesh size of 10% of the hole radius of the rock core plug holder. Then we increased the mesh size to 30% to enable faster simulation calculation while observing the mesh size effect on the results. We conclude that a mesh size of 30% is suitable, and there is no need to go to a smaller mesh size. The simulation results are in Figure 11. The cold side of the rock core plug holder, Figure 11, which has higher exposure to the laboratory environment, produces a localized effect along one side of the holder. This critical finding tells us that the NMR signal is mainly affected by the heating of the flooding fluid and less by the laboratory environment. However, we still need to consider the partial cooling effect that the simulation had identified. Therefore, we suggest a weighted average temperature value for correcting the NMR signal using Eq. 2 and 3. This finding would not be possible without simulation.

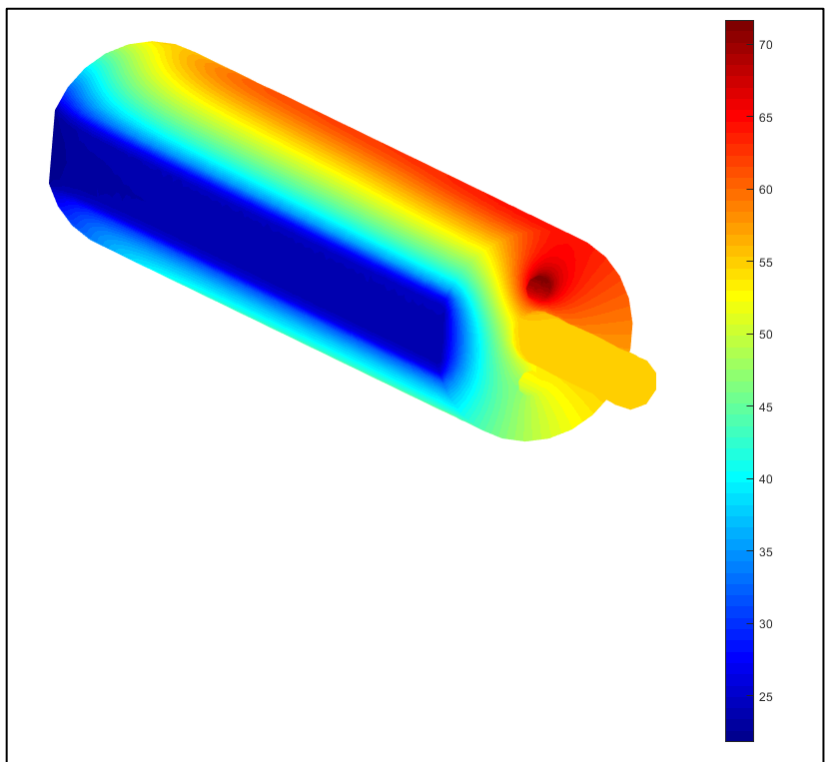

**Figure 11. Simulation results show that ambient cooling does not reach the rock core plug. However, it alters the holder temperature, an effect that we consider during NMR thermal correction (Temperature °C).**

## Conclusion
- NMR Signal shows adrift at elevated temperature; the higher the temperature, the higher the drift. This drift is due to the dipolar coupling effect caused by the higher ionic spinning freedom at a higher temperature.
- Experimentation approach results offered an essential insight into how temperature can change NMR signal while simultaneously providing insight into the rock core plug holder's heat transfer (cooling rate).
- The numerical simulation approach provided valuable assurance about the temperature distribution across the rock core flooding cylinder in 3D.
- The Th-CENS provides researchers with a thermal correction tool for future NMR experiments conducted under elevated temperatures.

**Authors Affiliation**
- Omar Alfarisi: Senior Engineer at the Department of Field Development of ADNOC Offshore, Adjunct Professor at the China University of Petroleum Qingdao, and Guest Lecturer at Khalifa University.
- Hongtao Zhang: Ph.D. Candidate at the Department of Mechanical Engineering of Khalifa University.
- Djamel Ouzzane: Senior Specialist at the Department of Upstream development of ADNOC
- Hongxia Li: Research Scientist at the Department of Mechanical Engineering of Khalifa University.
- Aikifa Raza: Research Scientist at the Department of Mechanical Engineering of Khalifa University.

**Remarks**
- This paper version is a preprint for submission at the osf.io library.
- This paper version was last updated on Jan 20th, 2021.

**Acknowledgment**

The authors would like to thank ADNOC, ADNOC Offshore, and Khalifa University for their support in progressing this research work. The authors would like to acknowledge the extraordinary and relentless support expressed by Professor TieJun Zhang for conducting this research work.